\documentclass[a4paper,11pt]{article}

\usepackage{contribution}

\usepackage{epstopdf}

\newcommand{\weblink}[2][]{%
    \ifthenelse{\equal{#1}{}}%
    {\textnormal{\url{#2}}}%
    {\textnormal{\href{#2}{#1}}}%
}

\newcommand{\acknowledgements}[1]{%
  \bigskip\bigskip
  \textsf{\textbf{\Large Acknowledgements}} \\[2ex]
  {#1}
  \bigskip
}

\def\beq{\begin{equation}}
\def\eeq#1{\label{#1}\end{equation}}
\def\eeqn{\end{equation}}

\def\beqa{\begin{eqnarray}}
\def\eeqa#1{\label{#1}\end{eqnarray}}
\def\eeqan{\end{eqnarray}}

\let\bar=\overbar

\def\Dslash{\not{\hbox{\kern-4pt $D$}}}
\def\dslash{\not{\hbox{\kern-2pt $\del$}}}

\def\msb{{\bar{\ssstyle M \kern -1pt S}}}

\newcommand{\contribution}[7][]{%
  \clearpage
  \thispagestyle{plain}
  \ifthenelse{\equal{#1}{}}
  {\hypersetup{pdftitle={#2}}}
  {\hypersetup{pdftitle={#1}}}
  \hypersetup{pdfauthor={{#3} {#4}}}
  {\centering\normalfont\LARGE\bfseries\sffamily #2 \par\nobreak}
  \lhead{}
  \chead{%
    \textit{\footnotesize XIV International Conference on Hadron Spectroscopy
      (\weblink[\textit{hadron2011}]{http://www.hadron2011.de}), 13-17 June 2011, Munich, Germany}%
  }
  \rhead{}
  \bigskip
  \begin{center}
    {#3} {#4}\ifthenelse{\equal{#6}{}}{}{\footnote{\weblink[#6]{mailto:#6}}}
    \ifthenelse{\equal{#7}{}}{}{#7} \\
    \textit{#5}
  \end{center}
  \bigskip
}

\renewcommand{\abstract}[1]{%
  \begin{center}
    \begin{minipage}{0.85\textwidth}
      \begin{footnotesize}
        #1
      \end{footnotesize}
    \end{minipage}
  \end{center}
  \bigskip
}

\begin{document}

%
%
%
%
%
{  


%

\contribution[Recent results from heavy-ions at LHC and RHIC]  
{Recent experimental results from the relativistic heavy-ion collisions at LHC and RHIC}  
{Ilya}{Selyuzhenkov}
{Research Division and ExtreMe Matter Institute EMMI,\\
GSI Helmholtzzentrum f\"ur Schwerionenforschung\\
Planckstra\ss e 1 64291 Darmstadt, Germany}  
{ilya.selyuzhenkov@gmail.com}  
{}  
%

\abstract{%
A new era has started in the field of relativistic heavy-ion physics with
lead beams delivered by the Large Hadron Collider (LHC) in November 2010.
In this proceedings I highlight the main results from experimental measurements
with Pb--Pb collisions at the incident energy of
2.76~TeV/nucleon recorded by the LHC experiments.
Recent experimental developments from the Relativistic Heavy Ion Collider (RHIC)
at the GeV incident energy scale are also discussed.
All together LHC and RHIC measurements provide new insights on
the properties and features of the new hot and dense form
of matter created in the course of the relativistic heavy-ion collision.
}
%

\section{Introduction}

The main goal of the more than a ten years of operation
of the Relativistic Heavy Ion Collider (RHIC),
and of the heavy-ion program recently launch at the
Large Hadron Collider (LHC)
is to study the properties of the hot and dense matter,
the so called quark-gluon plasma (QGP),
which is believed to have existed a few microseconds after the big bang.
By colliding heavy nuclei at relativistic energies
we heat up the normal cold matter and transfer it from the hadronic phase
to fireball of deconfined quarks and gluons,
which allows us to probe the QGP properties in the laboratory.

Theoretically the time evolution of the system created in the course of
heavy-ion collisions is
described by a sequence of several stages.
It starts from the initial pre-equilibrium state when
hard parton scattering occurs and gluonic fields are formed.
The next stage is the formation and then expansion of a thermalized state of matter,
the quark-gluon plasma, which is conventionally described by
hydrodynamics.
Consequently, the quarks and gluons are coupled (hadronize) into hadrons,
which ends with the phase of chemical and then kinetic freeze out
(all interactions are ceased at this moment).
The only data which directly accessible for experimentalist
is the information on this last hadronic stage.
Evidently, it is not possible to constrain theoretical models which pretend
to describe the evolution of the heavy-ion collision
and identify the one which most precisely reflects the nature
of the collision with a single measurement, and
so it is necessary to study many experimental observables
to pin down the properties of the QGP.

In this proceedings I highlight
the new results from the ALICE, CMS, and ATLAS Collaborations
from the first heavy-ion run at LHC in November 2010,
as well as recent experimental developments by the STAR and PHENIX Collaborations at RHIC. 
I start with a discussion of global properties of the collisions at LHC energies,
such as charged hadrons multiplicity, particle yields, and a measurement of the
collision freeze out volume from the
Bose-Einstein correlations of identical pions.
I then briefly review the results for
anisotropic flow which reflect collectivity in particles production
and allows us to experimentally constrain the
possible initial conditions of the collision and the QGP viscosity.
Switching to hard probes, I present the main
results on the suppression of particle production at high transverse momenta in heavy-ion collision.
I continue the discussion with a few highlights from the beam energy scan program
at RHIC aimed to probe the properties of the phase boundary
and search for the critical point in the QCD phase diagram.
I complete my proceedings with a few remarks on
probes of local parity violation in the strong interaction
which shows the potential to go beyond the scope of the QGP physics
with the heavy-ion programs at RHIC and LHC.

\section{\label{Sec:Yields}Particle yields}

\begin{figure}[htb]
  \begin{center}
    \includegraphics[width=0.49\textwidth]{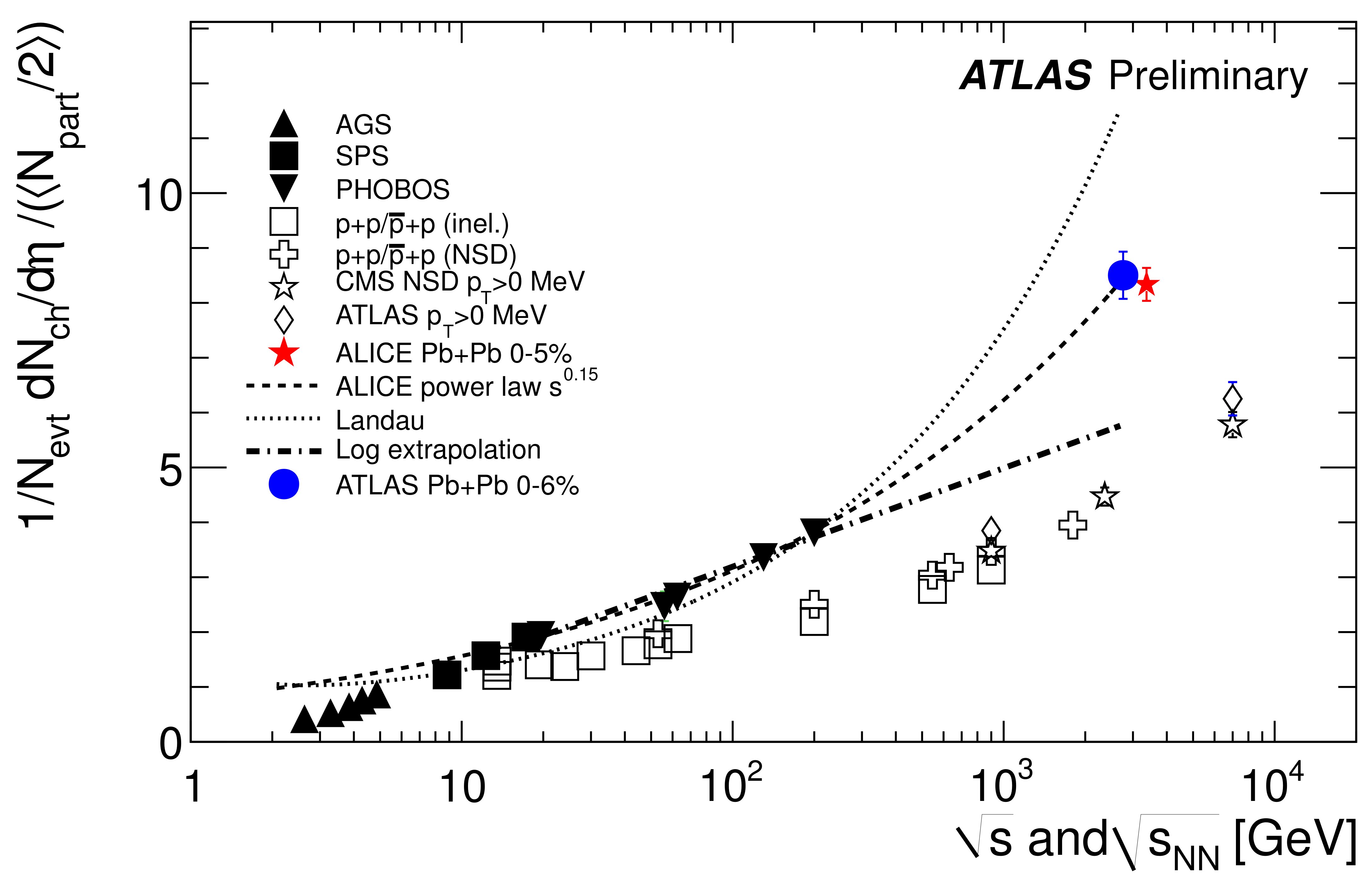}%
    ~~\includegraphics[width=0.51\textwidth]{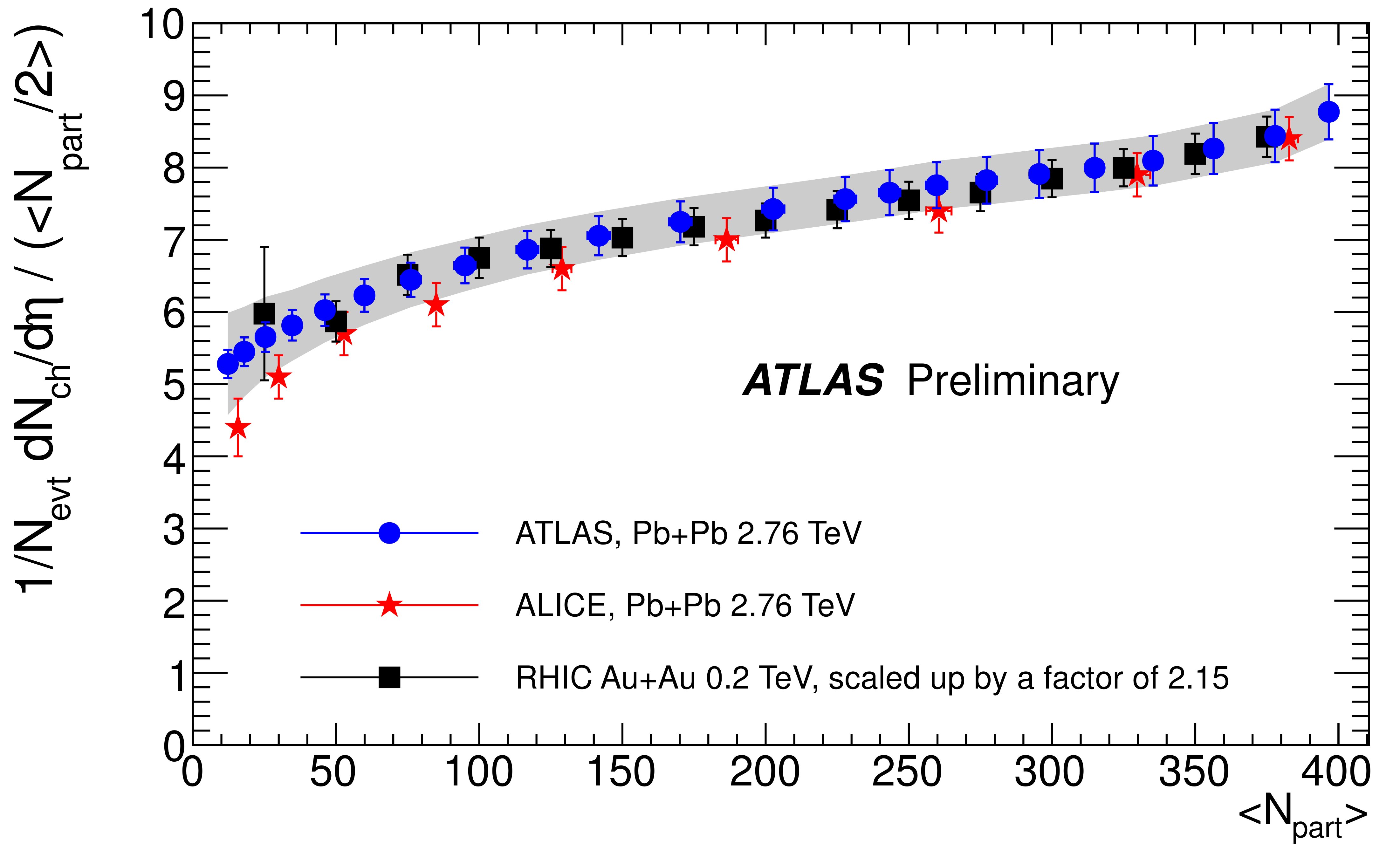}
    {\mbox{}\\\vspace{-0.7cm}
    \hspace{1cm}\mbox{~} \bf (a)
    \hspace{+6.7cm}\mbox{~} \bf (b)}
    {\mbox{}\\\vspace{-0.2cm}}
    \caption
    {
      (a)~Charged particle multiplicity per participant pair
      measured for Pb--Pb collisions by the ALICE and ATLAS Collaborations at LHC,
      and compared to results for proton-proton and
      heavy-ion collisions at lower energies.
      (b)~Centrality dependence of the multiplicity per participant
      pair measured by the ALICE and ATLAS Collaborations at the LHC, and the STAR Collaboration at RHIC.
      Figures taken from \cite{Steinberg:2011dj}.
    }
    \label{fig:ATLAS_ALICE_dndeta}
  \end{center}
\end{figure}
Figure \ref{fig:ATLAS_ALICE_dndeta}(a) shows a compilation of the results
for charged particle multiplicity density
measured for heavy-ion collisions at the LHC and lower energies at RHIC, SPS, and AGS,
as well as for proton-proton collisions.
The charged particle multiplicity density in central Pb--Pb collisions
at 2.76 TeV/nucleon is measured to be
${\rm d}N_{\rm ch}/{\rm d}\eta \approx 1600$,
what is larger by a factor of 2.15 than that at top RHIC energy.
Compared to pp collisions at the same energy 
the charge density is increased by a factor of 1.9.
The measured multiplicity and spectra correspond to an increase by 2.5-3 times in energy density from RHIC to LHC,
which for central Pb--Pb collisions at LHC is measured to
be $dE_{t}/d\eta \sim$ 2 TeV per unit of rapidity \cite{collaboration:2011ym}.
As it can be seen from Fig. \ref{fig:ATLAS_ALICE_dndeta}(b) the
shape of the charged particle production per participant pair versus centrality
is almost identical for  RHIC and LHC energies,
what may indicate that saturation effects do not significantly
change despite shifting toward smaller Feynman $x_f$ at LHC.

\begin{figure}[htb]
  \begin{center}
    \includegraphics[width=0.38\textwidth]{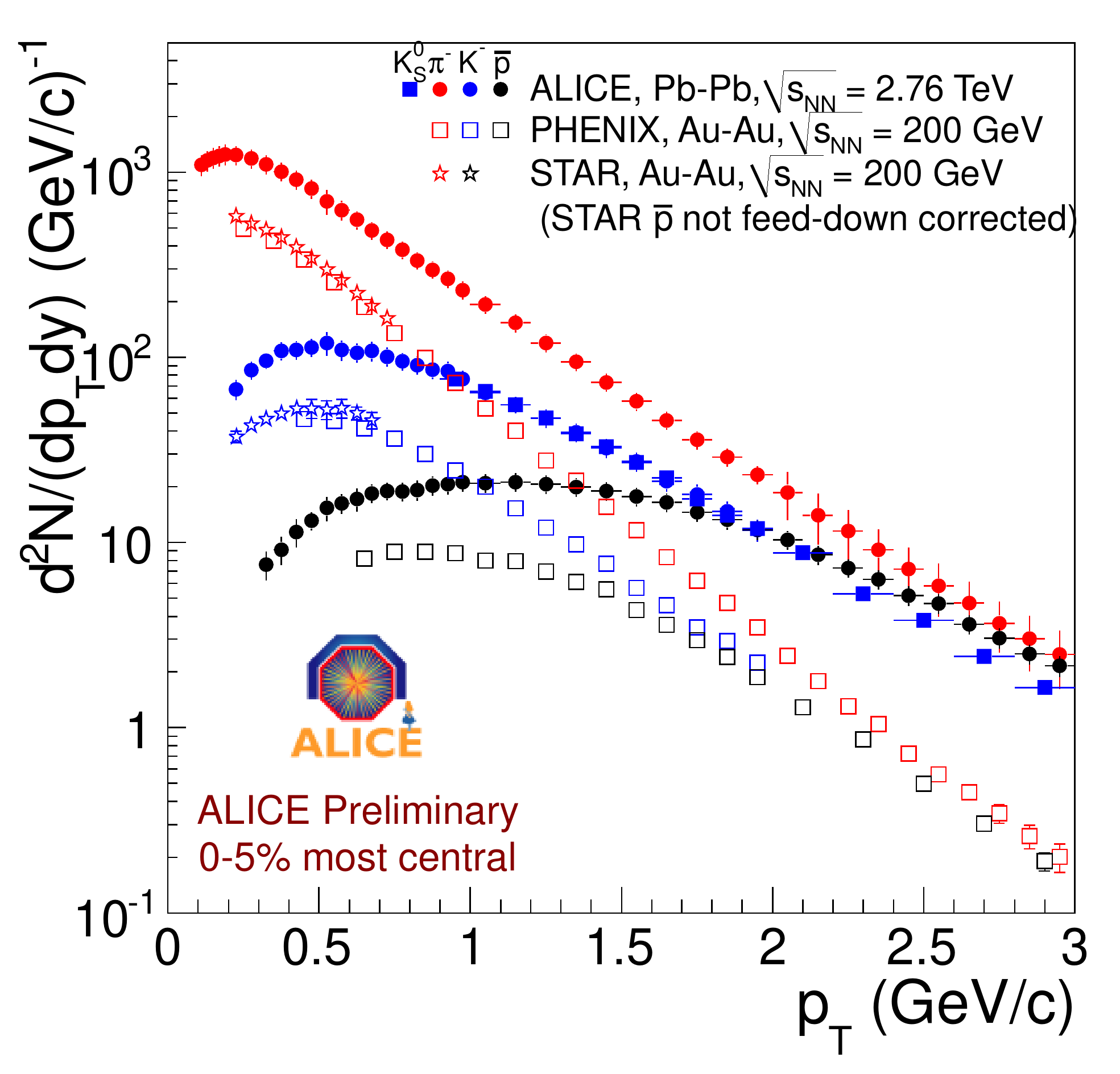}%
    \includegraphics[width=0.62\textwidth]{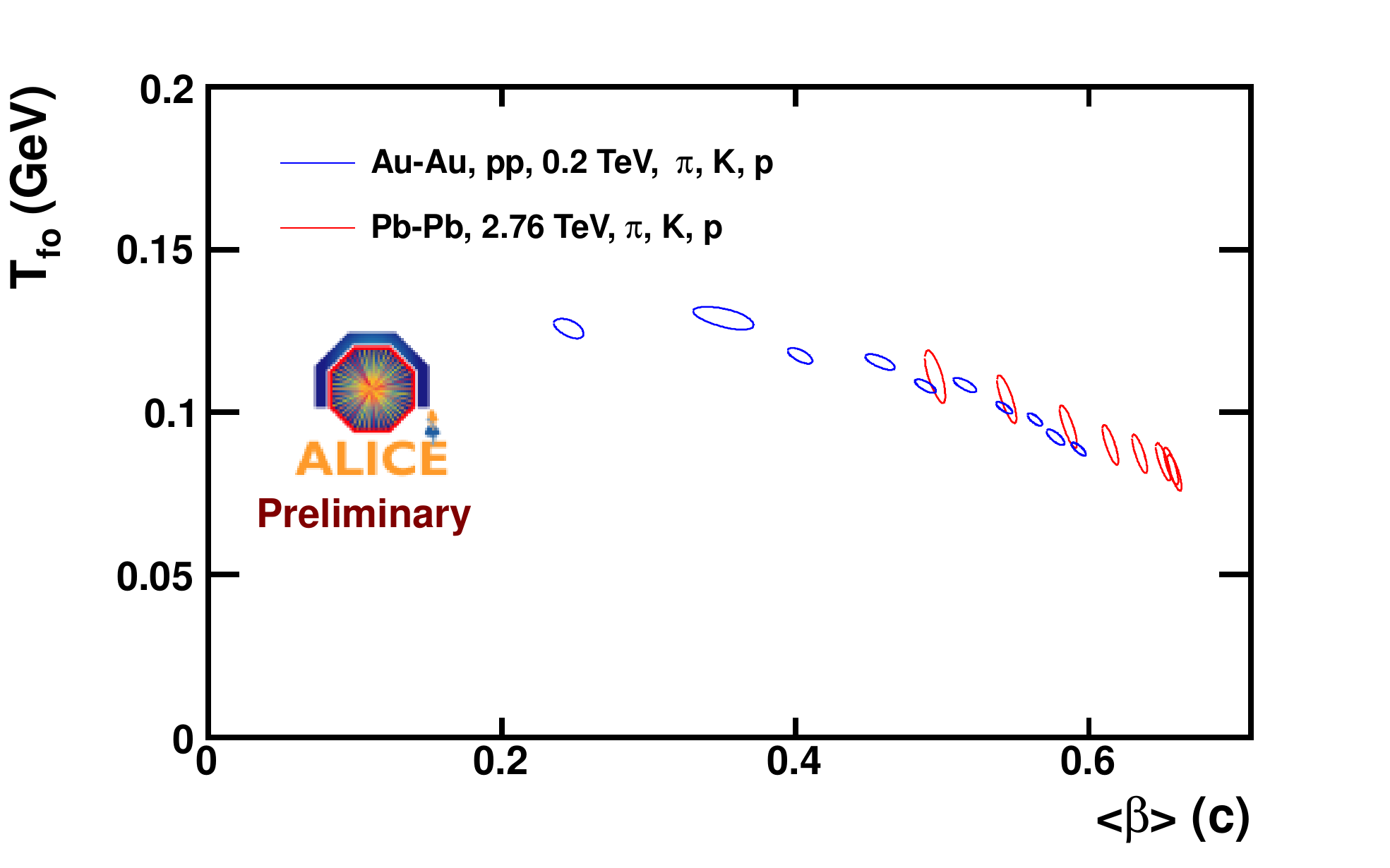}
    {\mbox{}\\\vspace{-0.7cm}
    \hspace{-0.9cm}\mbox{~} \bf (a)
    \hspace{+6.7cm}\mbox{~} \bf (b)}
    {\mbox{}\\\vspace{-0.2cm}}
    \caption
    {
      (a)~Identified charged particle spectra measured by the ALICE Collaboration for 
      heavy-ion collisions at the LHC in comparison with results for top RHIC energy.
      (b)~Freeze-out temperature, $T_{fo}$, and radial velocity, $\langle \beta_t \rangle$,
      extracted from the blast wave fits to the identified charged particle spectra
      measured at RHIC and LHC.
      Figures taken from \cite{Floris:2011ru}.
    }
    \label{fig:ALICE_Spectra}
  \end{center}
\end{figure}

Figure \ref{fig:ALICE_Spectra}(a) shows
the transverse momentum spectra of identified particle measured by the ALICE Collaboration
for the 0-5\% most central Pb--Pb collisions at 2.76 TeV/nucleon.
The spectra slopes change dramatically compared to RHIC data
(open symbols in Fig. \ref{fig:ALICE_Spectra}(a)), especially for protons.
This reflects a significantly stronger radial flow.
The radial flow velocity reaches about 60\% of the speed of light
with a simultaneous reduction of the kinetic freeze-out temperature down to 80 MeV
(Fig. \ref{fig:ALICE_Spectra}(b)).

\section{\label{Sec:HBT}Bose-Einstein correlations}
Figure \ref{fig:ALICE_hbt} shows the decoupling time and the size of the freeze-out (homogeneity) region
of the fireball created in Pb--Pb collisions at LHC 
and at lower energies.
\begin{figure}[htb]
  \begin{center}
    \includegraphics[width=0.48\textwidth]{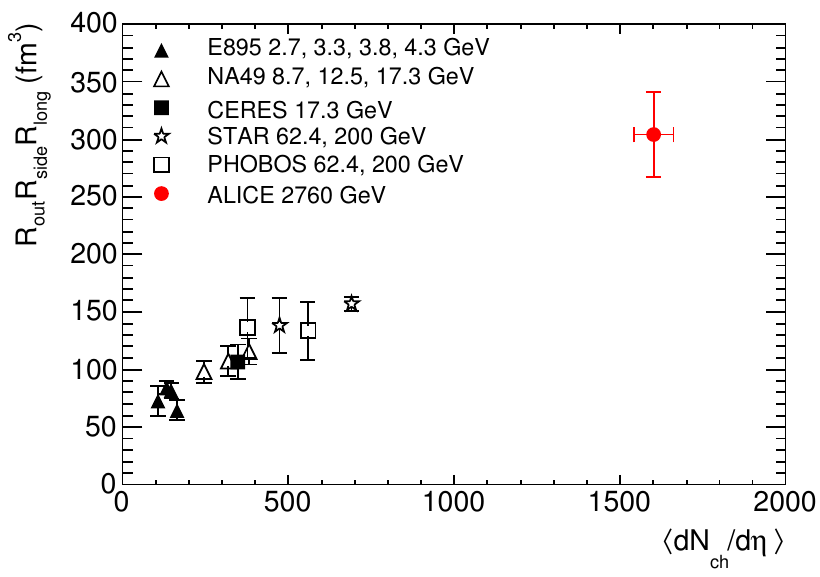}%
    \includegraphics[width=0.48\textwidth]{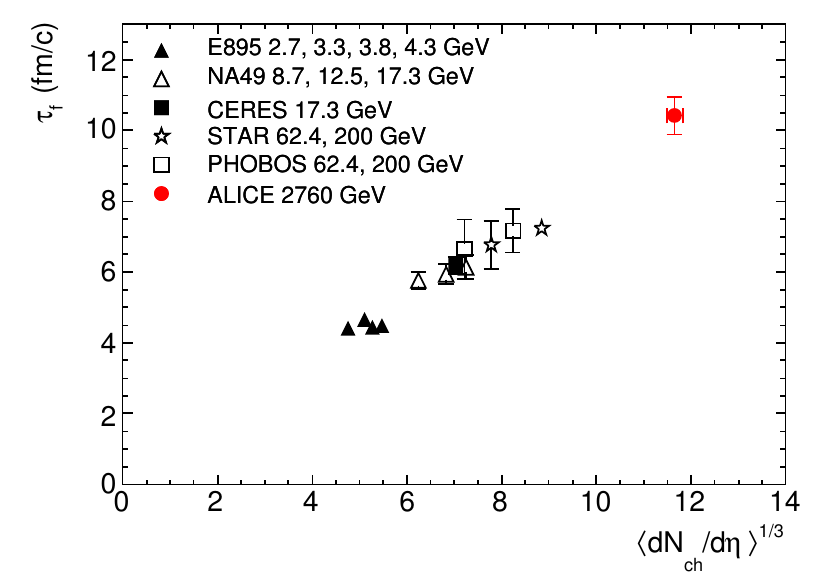}
    {\mbox{}\\\vspace{-0.5cm}
    \hspace{1cm}\mbox{~} \bf (a)
    \hspace{+6.7cm}\mbox{~} \bf (b)}
    {\mbox{}\\\vspace{-0.2cm}}
    \caption
    {
      (a)~The homogeneity volume (triple product of the pion HBT radii
      which is proportional to the system volume via $(2\pi)^{\frac{3}{2}}$ coefficient).
      (b)~The decoupling time  of the system created in the heavy-ion collision.
      Results are for Pb--Pb collisions at 2.76~TeV/nucleon measured by the ALICE Collaboration, and
      for central Au--Au and Pb--Pb collisions for lower AGS and RHIC energies.
      Figures taken from \cite{Aamodt:2011mr}.
    }
    \label{fig:ALICE_hbt}
  \end{center}
\end{figure}
The volume and the system lifetime are deduced from the Hanbury-Brown-Twiss (HBT)
momentum-space two-particle correlations of identical pions. 
The HBT homogeneity region,
which is connected to the HBT out- long- and side- radii
via $(2\pi)^{\frac{3}{2}}R_{\rm out}R_{\rm side}R_{\rm long}$,
increases by a factor 2 (Fig.~\ref{fig:ALICE_hbt}(a))
compared to the top RHIC energy of $0.2$ TeV/nucleon pair. 
The system lifetime also increases by more than 30\% (Fig.~\ref{fig:ALICE_hbt}(b)).
These trends are consistent with hydrodynamical model calculations for LHC energies
using parameters tuned to reproduce the RHIC data.

\section{\label{Sec:Flow}Anisotropic transverse flow}
Azimuthal anisotropic flow is a key observable indicating collectivity
among particles produced in relativistic heavy-ion collisions.
Figure \ref{fig:CMS_ATLAS_v2} shows the integrated (a) and $p_t$ differential (b) elliptic flow, $v_2$,
measured in Pb--Pb collisions at 2.76 TeV/nucleon.
\begin{figure}[htb]
  \begin{center}
    \includegraphics[width=0.38\textwidth]{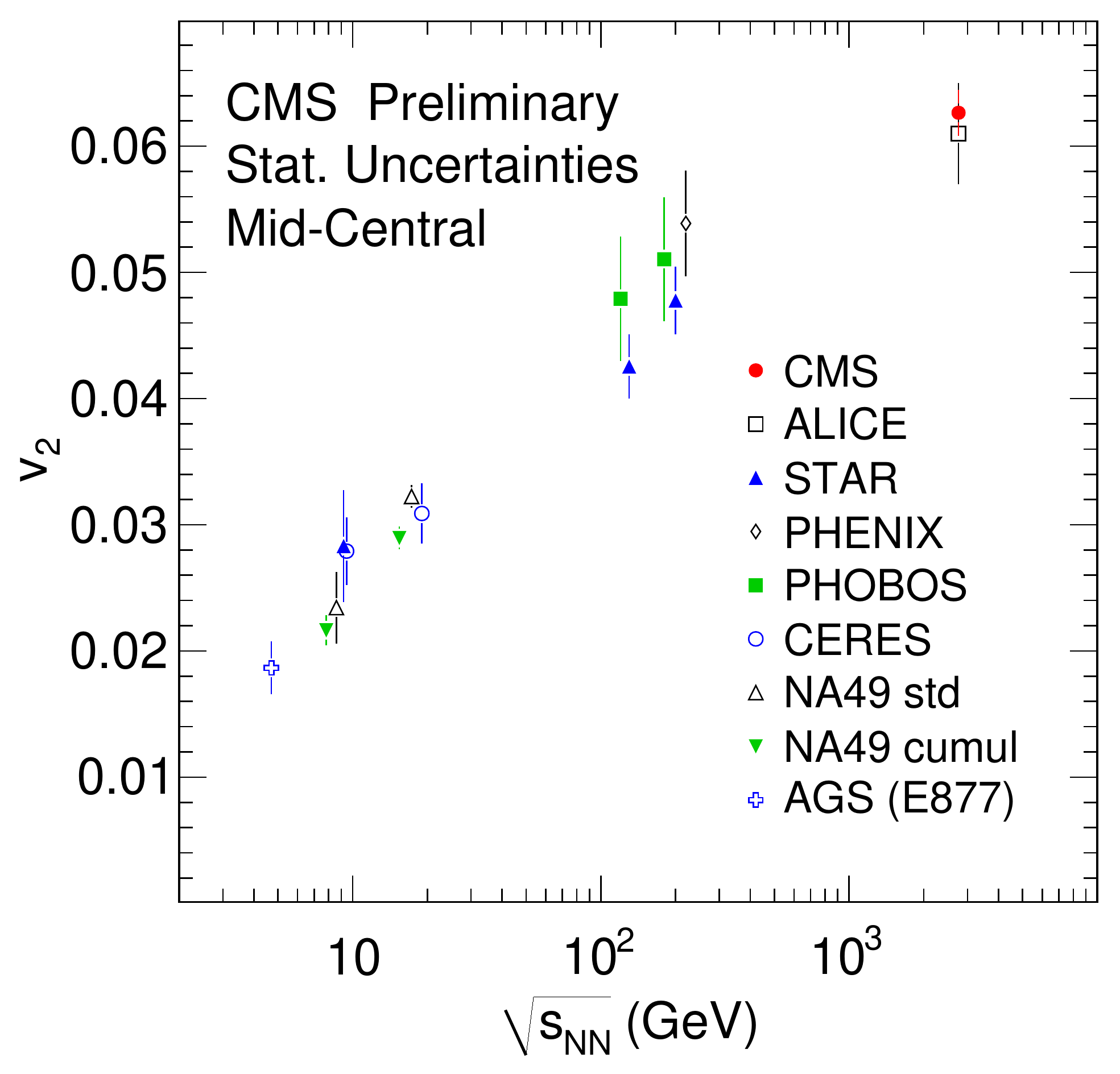}%
    ~~~~\includegraphics[width=0.52\textwidth]{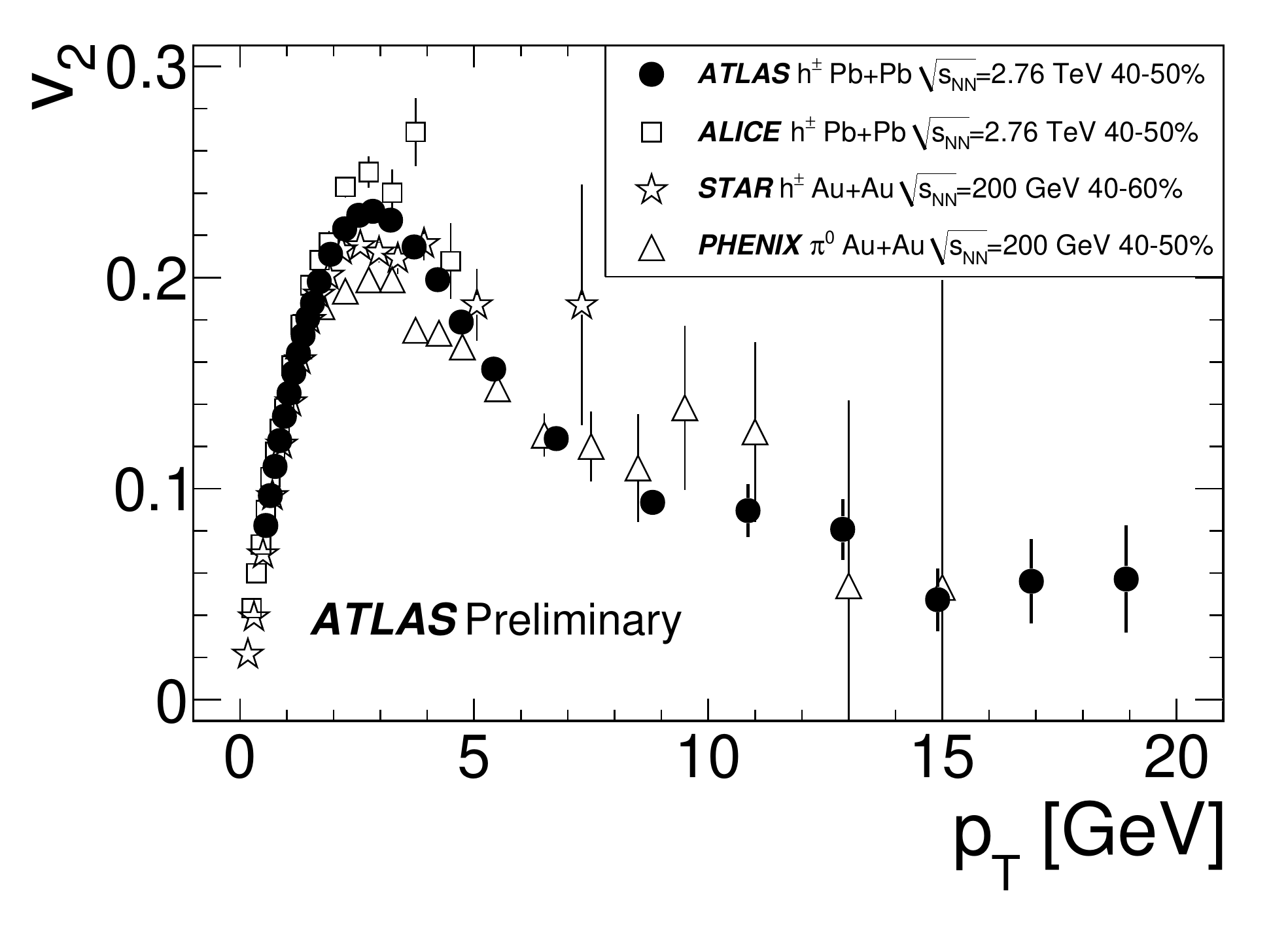}
    {\mbox{}\\\vspace{-0.2cm}
    \hspace{-0.3cm}\mbox{~} \bf (a)
    \hspace{+5.7cm}\mbox{~} \bf (b)}
    {\mbox{}\\\vspace{-0.2cm}}
    \caption
    {
      (a)~Integrated elliptic flow $v_2$ as a function of the collisions energy.
      (b)~$v_2$ as a function of transverse momentum for the 40-50\% centrality range
       measured in heavy-ion collisions at RHIC and LHC.
      Figure (a) taken from \cite{collaboration:2011ym}, figure (b) from \cite{Steinberg:2011dj}.
    }
    \label{fig:CMS_ATLAS_v2}
  \end{center}
\end{figure}
The immediate conclusion to be drawn from the 
comparison of $v_2$ results measured at LHC by the ALICE, ATLAS, and CMS Collaborations
to that at lower RHIC energies is that the 
integrated $v_2$ increases by 30\%, see Fig.~\ref{fig:CMS_ATLAS_v2}(a).
Figure~\ref{fig:CMS_ATLAS_v2}(b)
shows the differential flow results as a function of charged particle's transverse momentum, $p_{\rm t}$.
The results from RHIC and the LHC are similar
in both magnitude and
the shape of $p_{\rm t}$ dependence.
This behavior is a consequence of a stronger radial
flow at the LHC as was already discussed in Sec.~\ref{Sec:Yields} above.
The strong particle collectivity reflected by large $v_2$ at LHC
shows that the system created in heavy-ion collision at TeV energy scale
behaves as a strongly interacting, close to be a perfect fluid -
similar to the properties of the QGP observed at RHIC.
All this speaks towards applicability of the
hydrodynamic model description of the heavy-ion collisions at LHC energies.
\begin{figure}[htb]
  \begin{center}
    \includegraphics[width=0.48\textwidth]{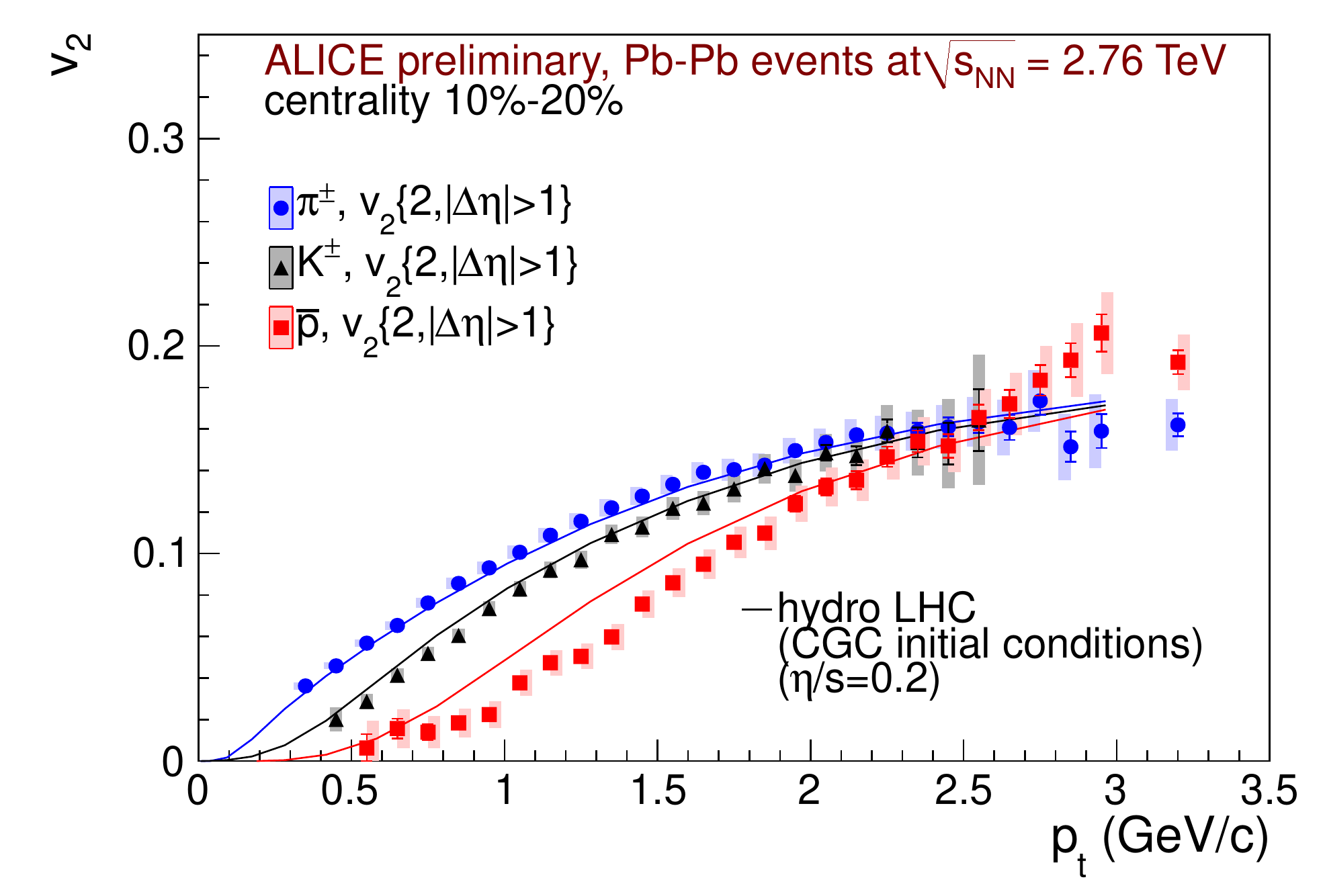}%
    \includegraphics[width=0.48\textwidth]{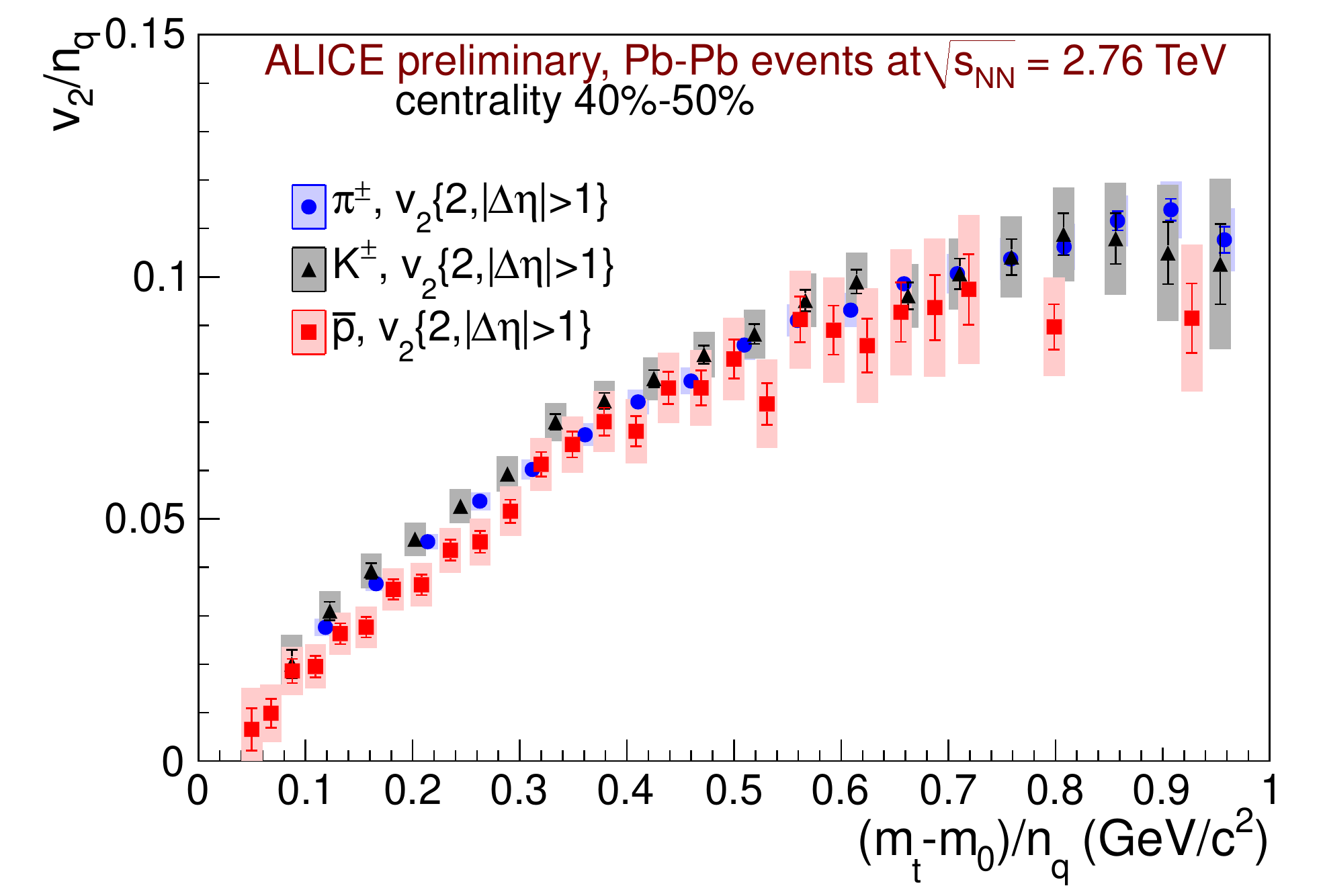}
    {\mbox{}\\\vspace{-0.5cm}
    \hspace{1cm}\mbox{~} \bf (a)
    \hspace{+6.7cm}\mbox{~} \bf (b)}
    {\mbox{}\\\vspace{-0.2cm}}
    \caption
    {
      (a)~Elliptic flow of pions, kaons and anti-protons
      vs. transverse momentum for the 10-20\% centrality range.
      The lines are hydrodynamical model calculations. 
      (b)~Elliptic flow versus transverse
      kinetic energy are both scaled with the number
      of constituent quarks for the 40-50\% centrality range.
      Figures taken from \cite{Collaboration:2011yb}.
    }
    \label{fig:ALICE_PIDv2}
  \end{center}
\end{figure}

Qualitatively the validity of the hydrodynamic description for LHC
and effects of stronger radial
flow can be tested with the anisotropic flow measurement
of identified particles and its dependence on the mass of different species.
Figure~\ref{fig:ALICE_PIDv2} shows the $p_t$ differential elliptic flow
of charged pions, kaons, and protons
measured by the ALICE Collaboration in Pb--Pb collisions at 2.76 TeV/nucleon.
The observed larger than at RHIC mass splitting of $v_2$ agrees well
with a picture of increased radial flow
and follows viscous hydrodynamic predictions (solid lines in Fig.~\ref{fig:ALICE_PIDv2}(a))
except for anti-protons in the most central collisions.
Anti-protons also fall out of the universal scaling with number of quarks
seen at RHIC energies.

One of the main highlights of recent anisotropic flow results
from both RHIC and LHC experiments
is establishing the connection
between the measured event anisotropy in the momentum space (anisotropic flow)
and the fluctuations of the energy density in the initial state of the heavy-ion collisions.
\begin{figure}[htb]
  \begin{center}
    \includegraphics[width=0.49\textwidth]{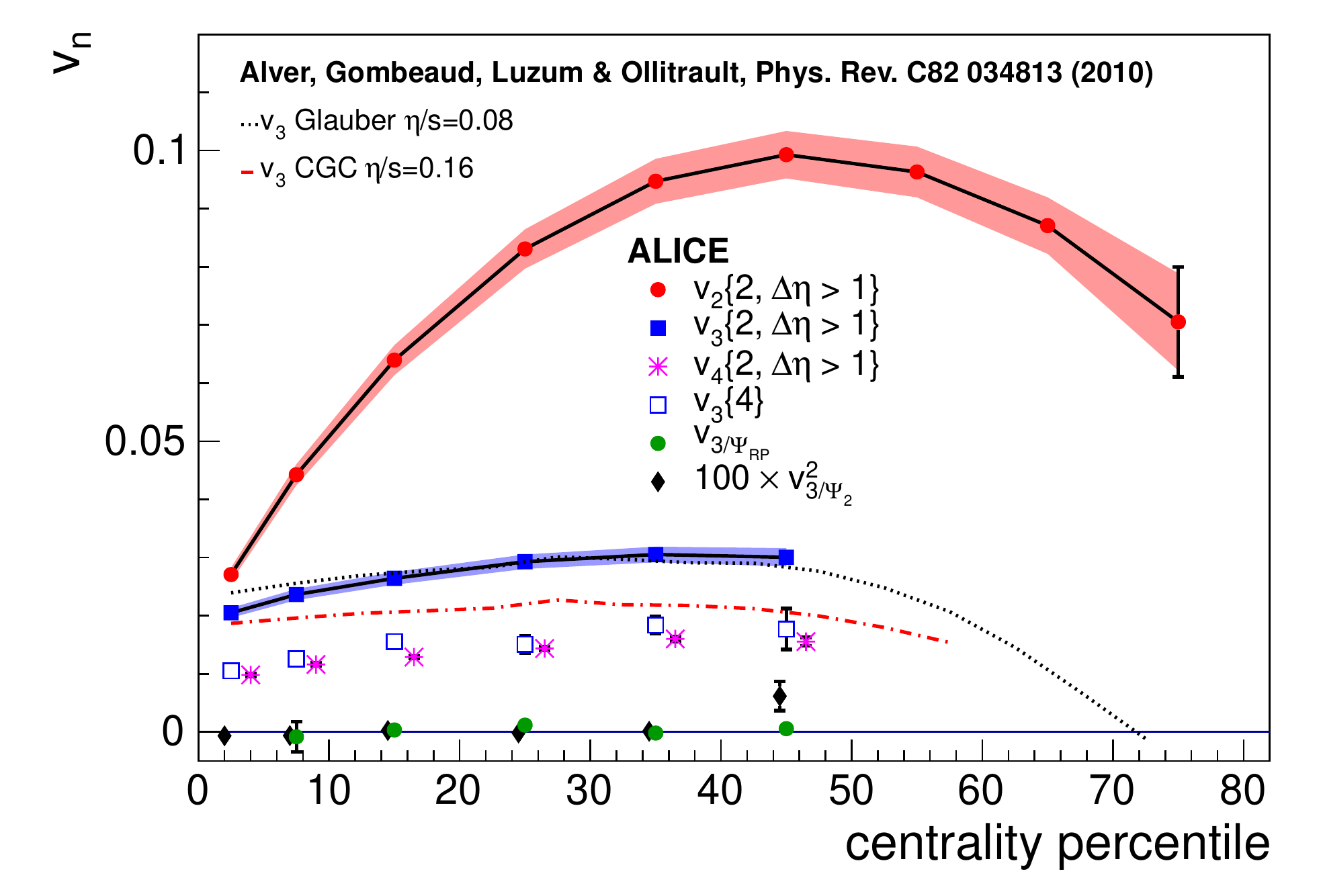}%
    \includegraphics[width=0.51\textwidth]{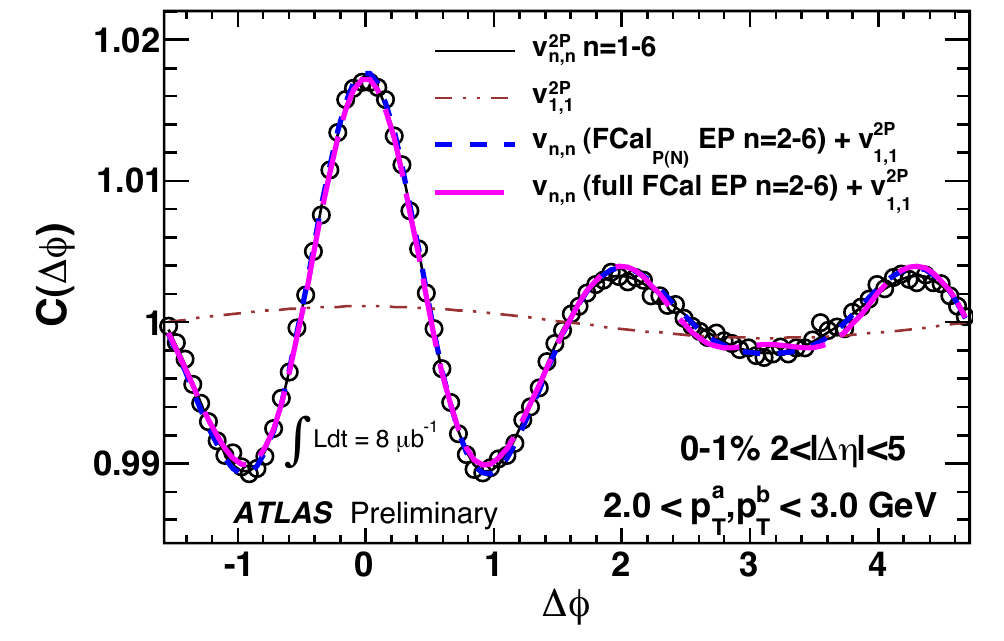}
    {\mbox{}\\\vspace{-0.7cm}
    \hspace{1cm}\mbox{~} \bf (a)
    \hspace{+6.7cm}\mbox{~} \bf (b)}
    {\mbox{}\\\vspace{-0.2cm}}
    \caption
    {
      (a)~Integrated elliptic ($v_2$), triangular ($v_3$), and quadrangular ($v_4$) flow
      measured for Pb--Pb collisions at 2.76 TeV/nucleon by the ALICE Collaboration.
      (b)~Two-particle correlation function measured for 1\% most central collisions by the ATLAS Collaboration.
      The measured 2-particle correlations are reproduced well by the combination
      of the Fourier coefficients from the anisotropic flow measurement
      (solid lines). Figure (a) taken from \cite{Collaboration:2011yb}, figure (b) from \cite{Steinberg:2011dj}.
    }
    \label{fig:ALICE_PHENIX_vN}
  \end{center}
\end{figure}
Figure \ref{fig:ALICE_PHENIX_vN}(a) 
shows higher order harmonics of anisotropic flow ($v_3$, $v_4$)
together with the largest $v_2$ component
for Pb--Pb collisions
at 2.76 TeV/nucleon measured by the ALICE Collaboration.
The geometrical origin of the $v_3$ component is established
by comparison of the vanishing triangular flow, $v_3$,
when measured with respect to the collision reaction plane
(green symbols in Fig. \ref{fig:ALICE_PHENIX_vN}(a)),
vs. non-zero $v_3$ measured with respect to the participant plane -
the plane determined by
the event-by-event fluctuating shape of the initial energy density (blue squares in Fig. \ref{fig:ALICE_PHENIX_vN}(a)).
Triangular and higher order harmonic flow also explains the double hump structure
seen originally at RHIC in the two-particle azimuthal correlations
and often referred to as the Mach Cone effect.
Figure~\ref{fig:ALICE_PHENIX_vN}(b) shows that
for the most central Pb--Pb collisions at 2.76 TeV/nucleon
the whole shape of the two particle azimuthal correlations
is driven by the interplay between various
anisotropic flow components, mainly $v_2$ and $v_3$.

\section{\label{Sec:Raa}Particle production at large transverse momenta}
Production of particles with very large transverse momentum,
$p_{\rm t}$, in heavy-ion collisions happens very early in the collision history
and therefore these particles have to propagate through the hot and dense medium
created in the collision.
Consequently, the modification of high $p_{\rm t}$ particle production
compared to the production without medium (e.g. in proton-proton collisions)
carries information about the medium properties such as the energy loss mechanism
and its dependence on the path length.
Quantitatively the modification of particle production 
is described by the nuclear modification factor, $R_{\rm AA}$,
which presents the ratio of the particle yields in heavy-ion collision
to that of the proton-proton collisions scaled
by the corresponding number of binary collisions.
\begin{figure}[htb]
  \begin{center}
    \includegraphics[width=0.38\textwidth]{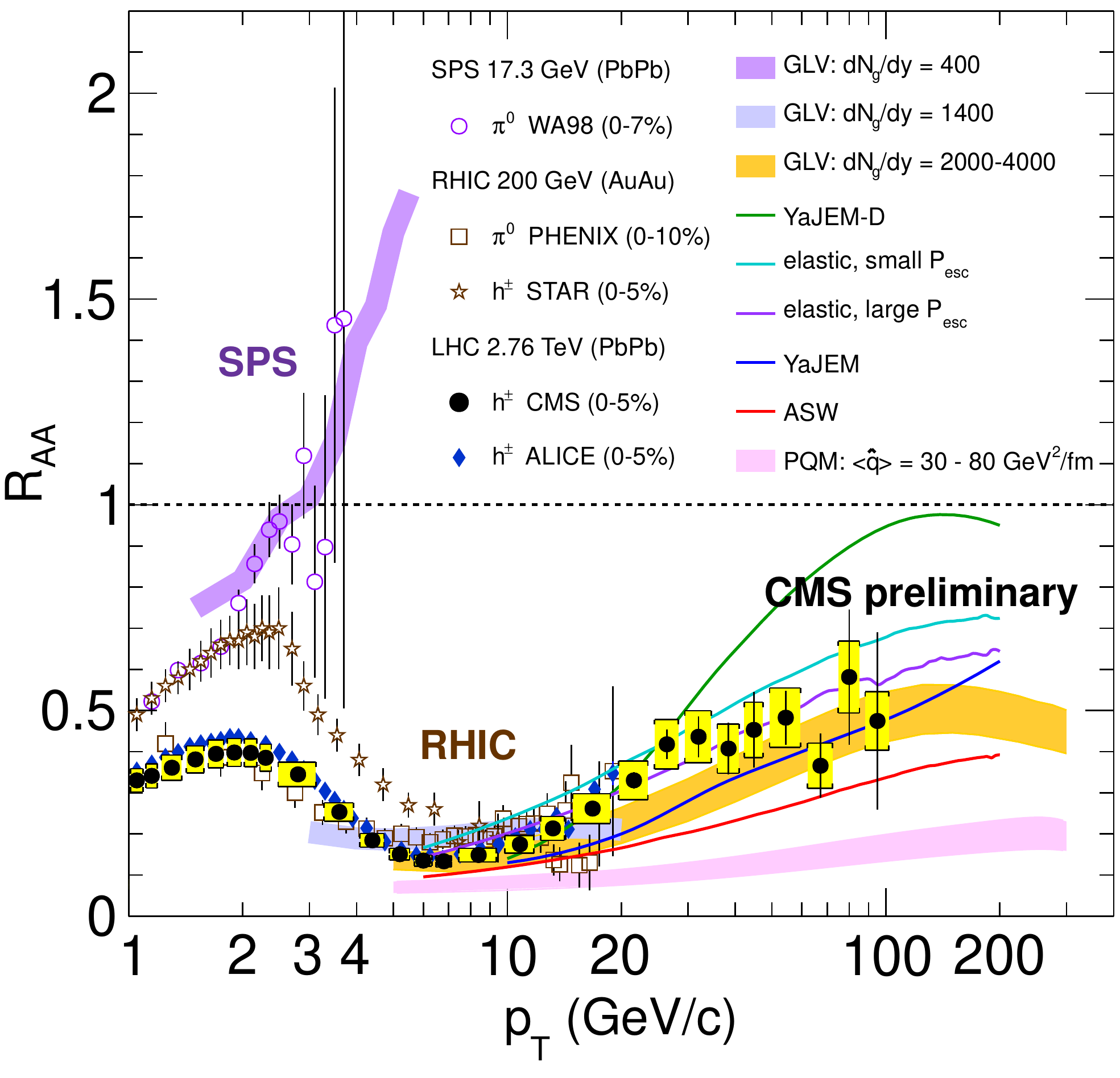}%
    \includegraphics[width=0.56\textwidth]{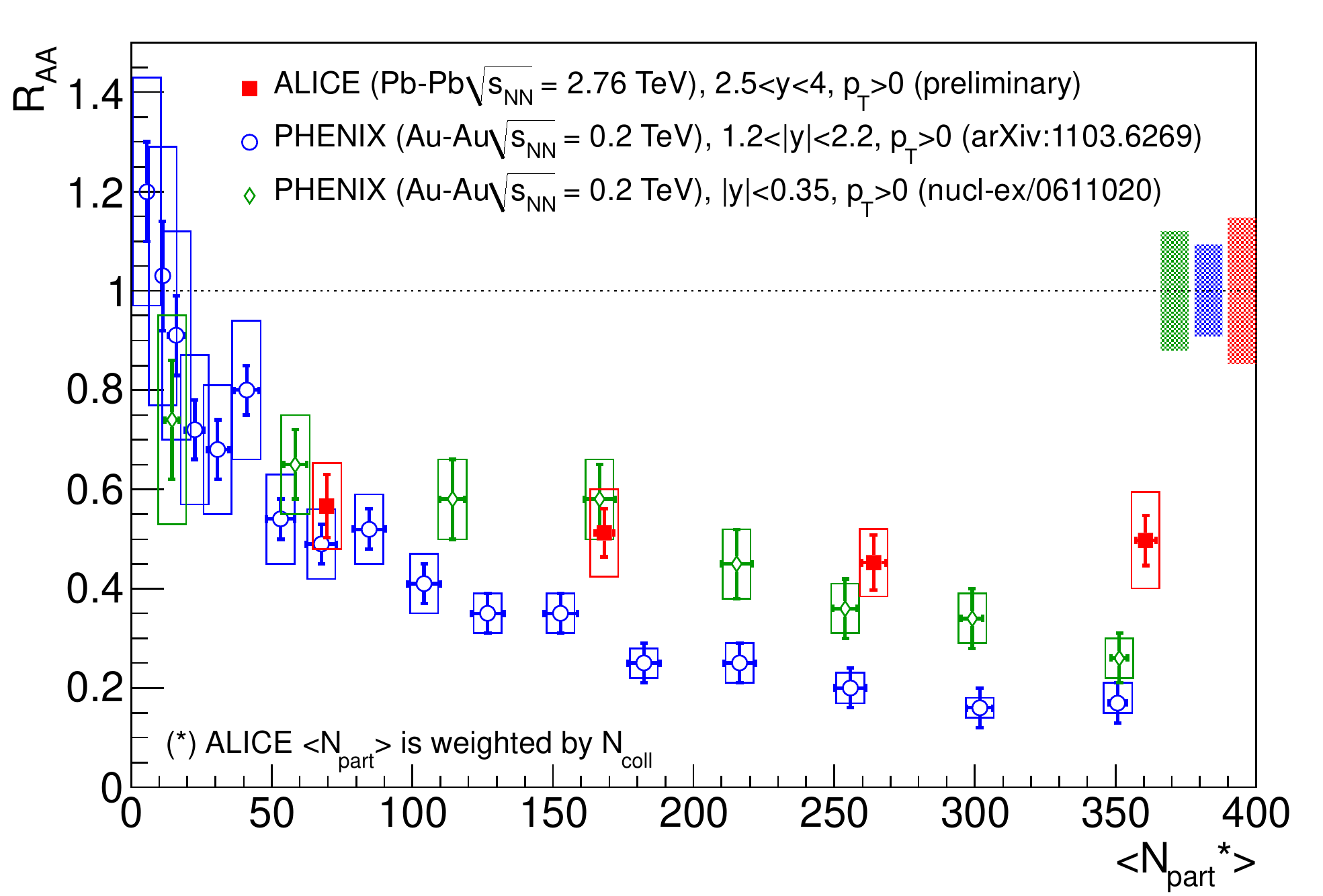}%
    {\mbox{}\\\vspace{-0.2cm}
    \hspace{-.5cm}\mbox{~} \bf (a)
    \hspace{+6.2cm}\mbox{~} \bf (b)}
    {\mbox{}\\\vspace{-0.2cm}}
    \caption
    {
      (a)~Nuclear modification factor, $R_{\rm AA}$, as a function of transverse momentum
      for neutral pions and charged hadrons for the most central heavy-ion collisions.
      (b)~$R_{\rm AA}$ of J/$\psi$ as a function of the number of participants
      measured at RHIC and LHC. Figure (a) taken from \cite{collaboration:2011ym},
      figure (b) from \cite{MartinezGarcia:2011nf}.
    }
    \label{fig:CMS_ALICE_Raa}
  \end{center}
\end{figure}
Figure~\ref{fig:CMS_ALICE_Raa}(a) shows the charged particle nuclear modification factor, $R_{\rm AA}$,
for Pb--Pb collisions at 2.76 TeV/nucleon measured by the ALICE and CMS Collaborations
compared to results for charged and identified neutral pions from RHIC and SPS experiments.
The deviation of $R_{\rm AA}$ from unity reflects the effect of medium modification,
and reveals strong suppression ($R_{\rm AA}<<1$) of particle
production in heavy-ion collisions compared to that in proton-proton interactions.
As a function of transverse momentum, $\rm R_{\rm AA}$ shows a minimum around 5-7 GeV/$c$,
and then rises significantly towards higher transverse momentum
but even at $p_{\rm t}\sim 100$ GeV/$c$ the
particle production is largely suppressed ($R_{\rm AA}\sim 0.5$).
Compared to different model calculations (color lines in Fig.~\ref{fig:CMS_ALICE_Raa}(a))
these new results provide strong constrains on models with different parton energy loss.
Figure~\ref{fig:CMS_ALICE_Raa}(b) shows that even heavy quark (J/$\psi$)
production is strongly suppressed at RHIC and LHC,
though at LHC the suppression is reduced in accord
with the expectations from the statistical model \cite{Andronic:2011yq}.

\begin{figure}[htb]
  \begin{center}
    \includegraphics[width=0.4\textwidth]{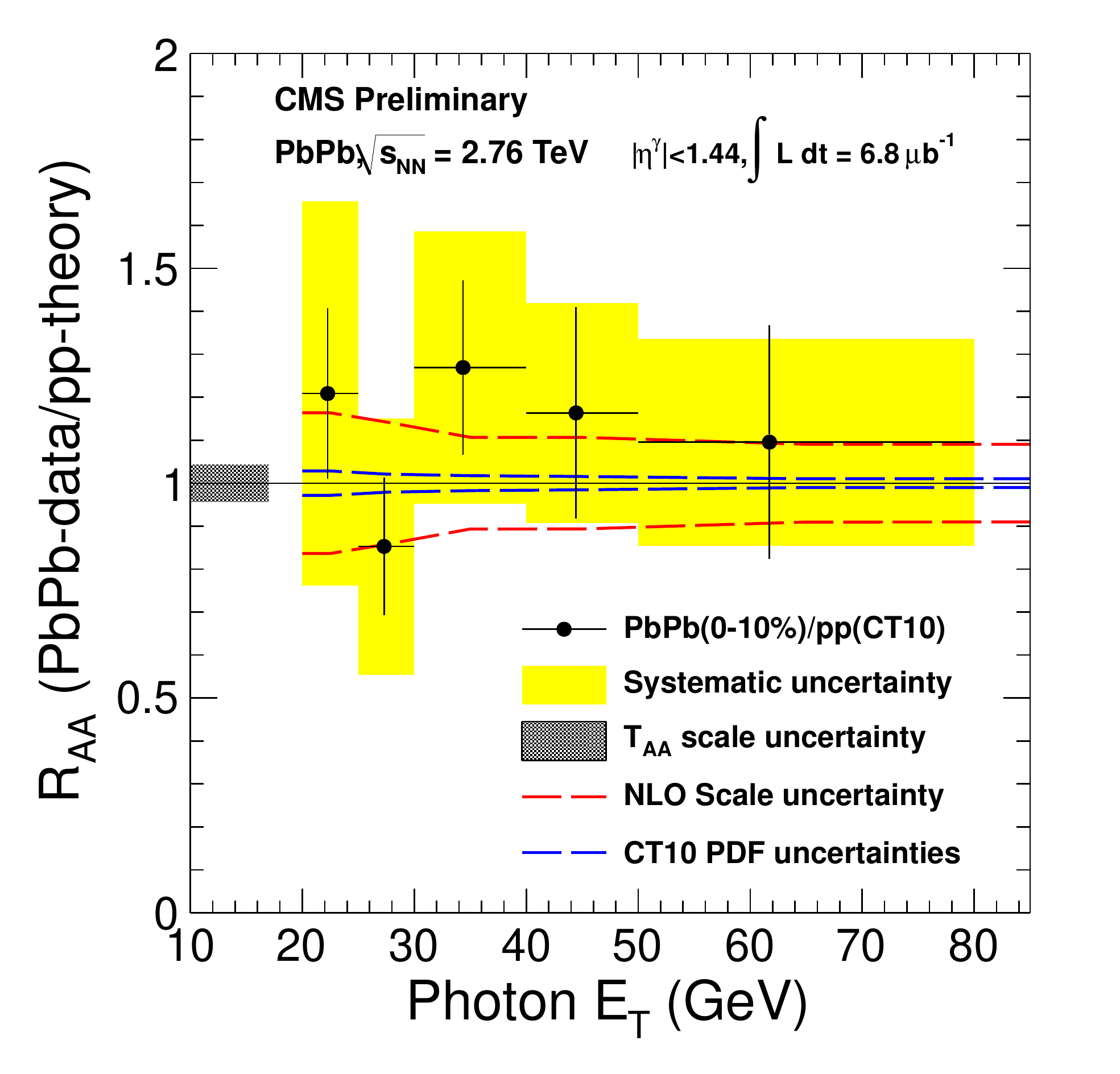}%
    \includegraphics[width=0.535\textwidth]{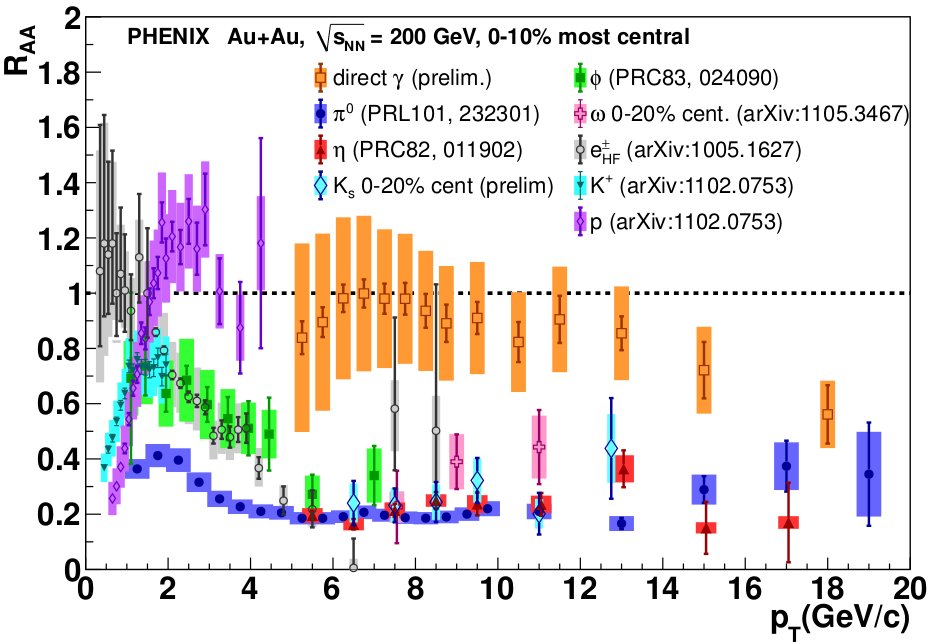}
    {\mbox{}\\\vspace{-0.2cm}
    \hspace{0cm}\mbox{~} \bf (a)
    \hspace{+6.7cm}\mbox{~} \bf (b)}
    {\mbox{}\\\vspace{-0.2cm}}
    \caption
    {
      (a)~Nuclear modification factor, $R_{\rm AA}$, of isolated photons as a function of
      transverse momentum for 0-10\% central events measured by the CMS Collaboration. 
      (b)~Results by the PHENIX Collaboration
      for $R_{\rm AA}$ of several mesons and direct photons for the 0-10\% central Au--Au collisions.
      Figure (a) taken from \cite{collaboration:2011ym}, figure (b) from \cite{Sharma:2011qm}.
    }
    \label{fig:CMS_PHENIX_Raa}
  \end{center}
\end{figure}
Important probes of the nuclear parton densities created in heavy-ion collision
are the colorless objects such as prompt photon and $Z$ boson,
since they are produced directly from hard parton interactions
and propagate through the medium of quarks and gluons without modification.
Figure~\ref{fig:CMS_PHENIX_Raa}(a) shows the $R_{\rm AA}$ of direct photons
as a function of the photon transverse energy for the 
most central Pb--Pb collisions measured by the CMS Collaboration.
Photon $R_{AA}$ measured at LHC is consistent with unity (no modification),
which is similar to the recent results for prompt photons by the PHENIX Collaboration
(orange points in Fig.~\ref{fig:CMS_PHENIX_Raa}(b)).
The $R_{\rm AA}$ of another colorless probe, the $Z$ boson,
is also measured by the CMS Collaboration
and found to be consistent with no medium modification:
$R_{\rm AA}^{(Z)} = 1.2\pm 0.29$(stat.)$\pm 0.16$(syst.) \cite{Lee:2011cs}.

\section{\label{Sec:RHIC_BES}RHIC beam energy scan and the search for the critical point}

Another frontier of the heavy-ion physics program,
which complements the study of the QGP properties at RHIC and LHC,
is to determine the nature of the phase transition
between confined (hadrons) and deconfined (quark-gluon plasma) matter
with a search for the critical point on the QCD phase diagram.
These two objectives are the main goals of the
Beam Energy Scan program at the RHIC facility.
The features of the phase transition and
proximity of the critical point
can be studied by looking at irregular changes in the
degrees of freedom of the system created in heavy-ion collisions.
Experimentally this should be reflected in non-monotonic behavior of the
sensitive physics observables.
Examples of sensitive observables are particle collectivity
such as anisotropic flow or HBT correlations,
or fluctuations in the system (e.g.
fluctuations of the conserved quantities such as baryon number or strangeness).

\begin{figure}[htb]
  \begin{center}
    \includegraphics[width=0.495\textwidth]{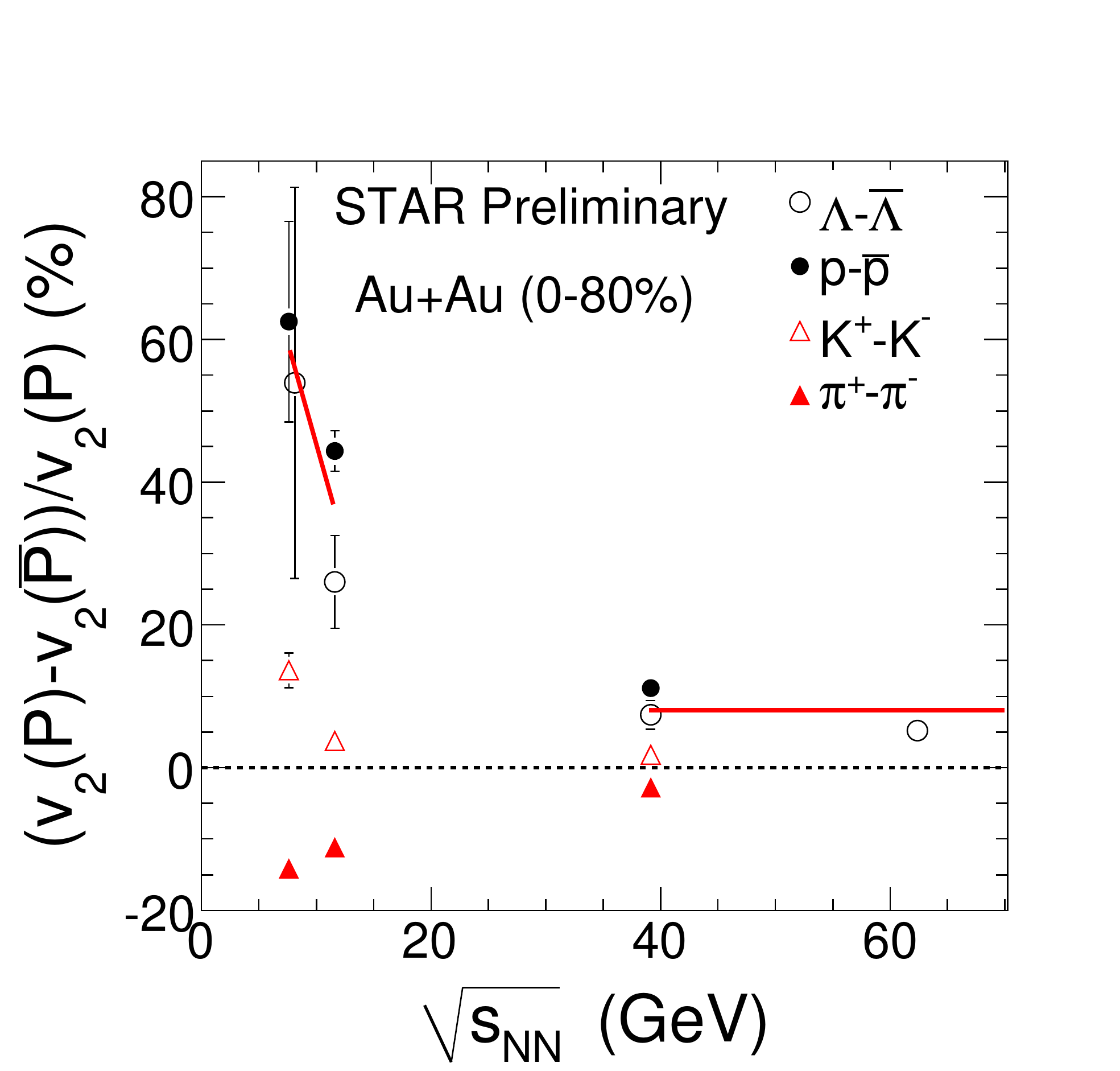}%
    \includegraphics[width=0.505\textwidth]{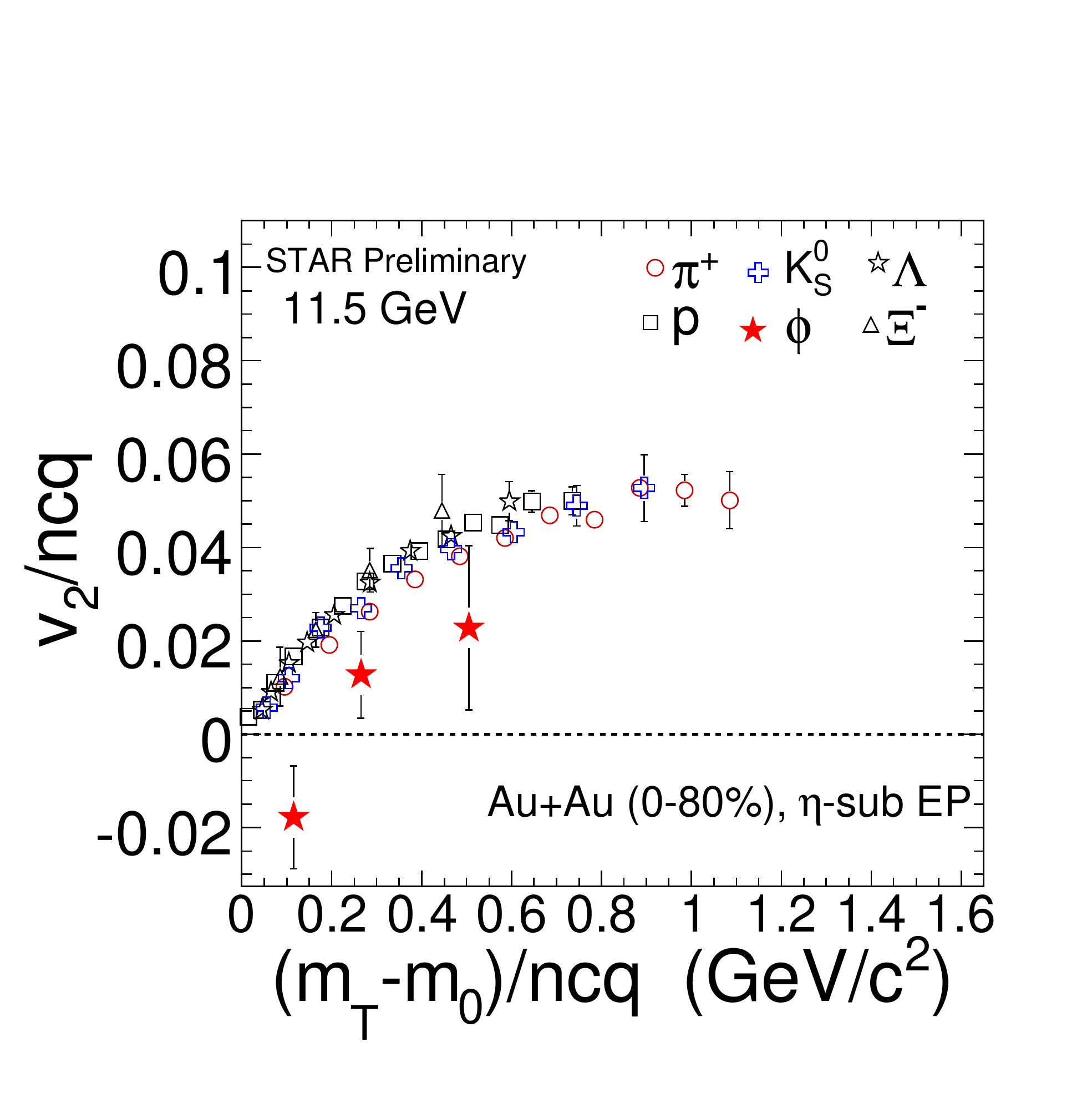}
    {\mbox{}\\\vspace{-0.5cm}
    \hspace{1cm}\mbox{~} \bf (a)
    \hspace{+6.7cm}\mbox{~} \bf (b)}
    {\mbox{}\\\vspace{-0.2cm}}
    \caption
    {
      (a)~Fractional difference between elliptic flow of particles and anti-particles
      for 0-80\% Au--Au collisions plotted vs. collision energy.
      (b)~Elliptic flow of identified particles for Au--Au
      collisions at 11.5 GeV/nucleon versus transverse
      kinetic energy both scaled by the number of constituent quarks.
      Figures taken from \cite{Mohanty:2011nm}.
    }
    \label{fig:STAR_bes}
  \end{center}
\end{figure}

Figure \ref{fig:STAR_bes}(a) shows the relative variation in the identified
particle and anti-particle elliptic flow measured by the STAR Collaboration
for different collision energies ranging from 11.5 GeV/nucleon
up to the top RHIC energy of 200 GeV/nucleon.
With decreasing collision energy the baryon and anti-baryon
elliptic flow difference increases dramatically compared to that for mesons.
Another change in the elliptic flow pattern at lower energies is the breaking
of the constituent quark scaling for the elliptic flow
which seems to hold at top RHIC energy.
Figure~\ref{fig:STAR_bes}(b) shows number of constituent quark scaled elliptic flow
of identified particles for Au--Au collisions at 11.5~GeV/nucleon.
Despite the large statistical errors, results for the
$\phi$-meson (which carries the information about the strange quark production)
deviates significantly from the overall scaling of other particles,
which probably indicate the breaking of the quark collectivity at lower energies
and change in the degrees of freedom in the system.

\begin{figure}[htb]
  \begin{center}
    \includegraphics[width=.42\textwidth]{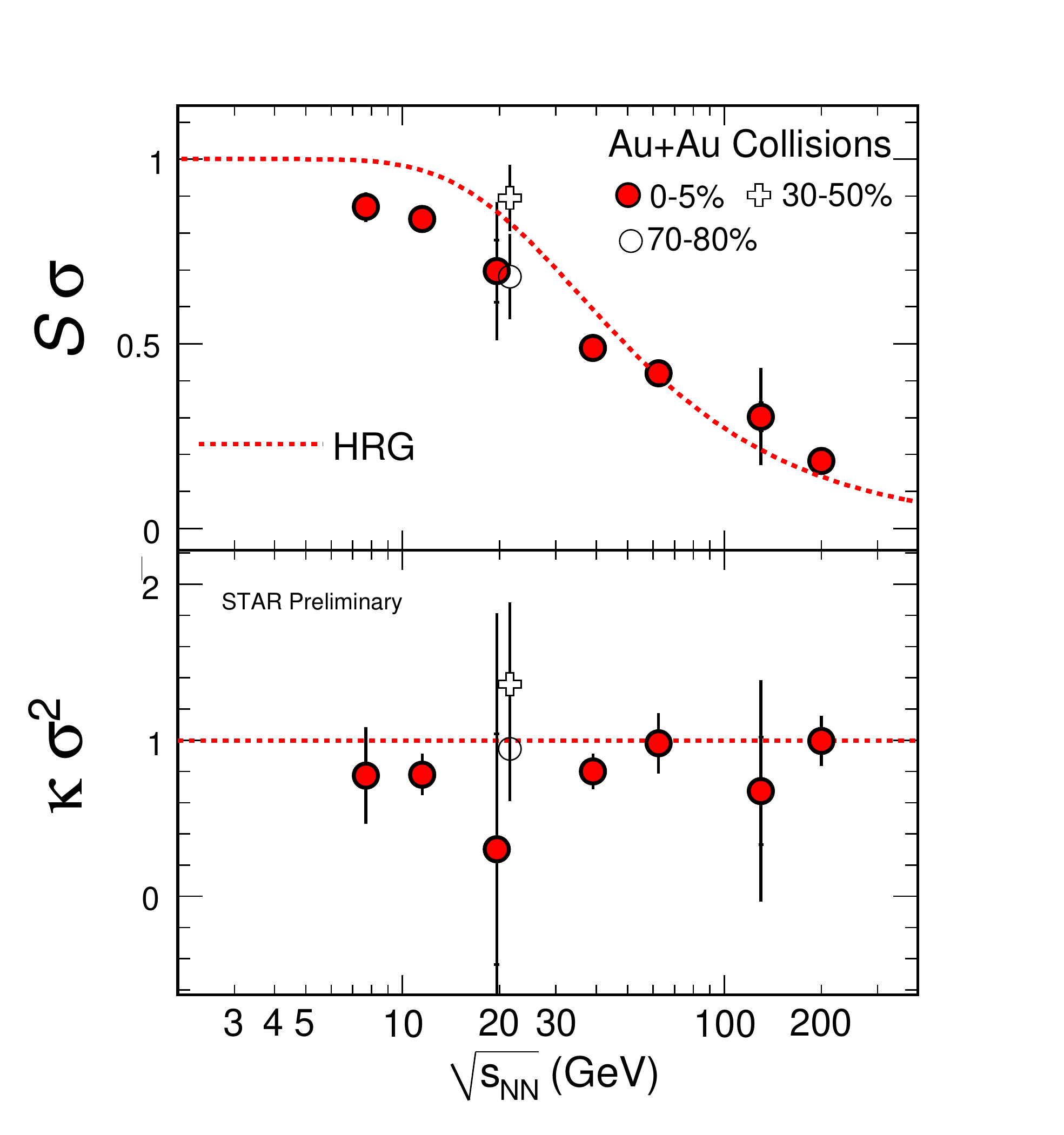}%
    \includegraphics[width=.57\textwidth]{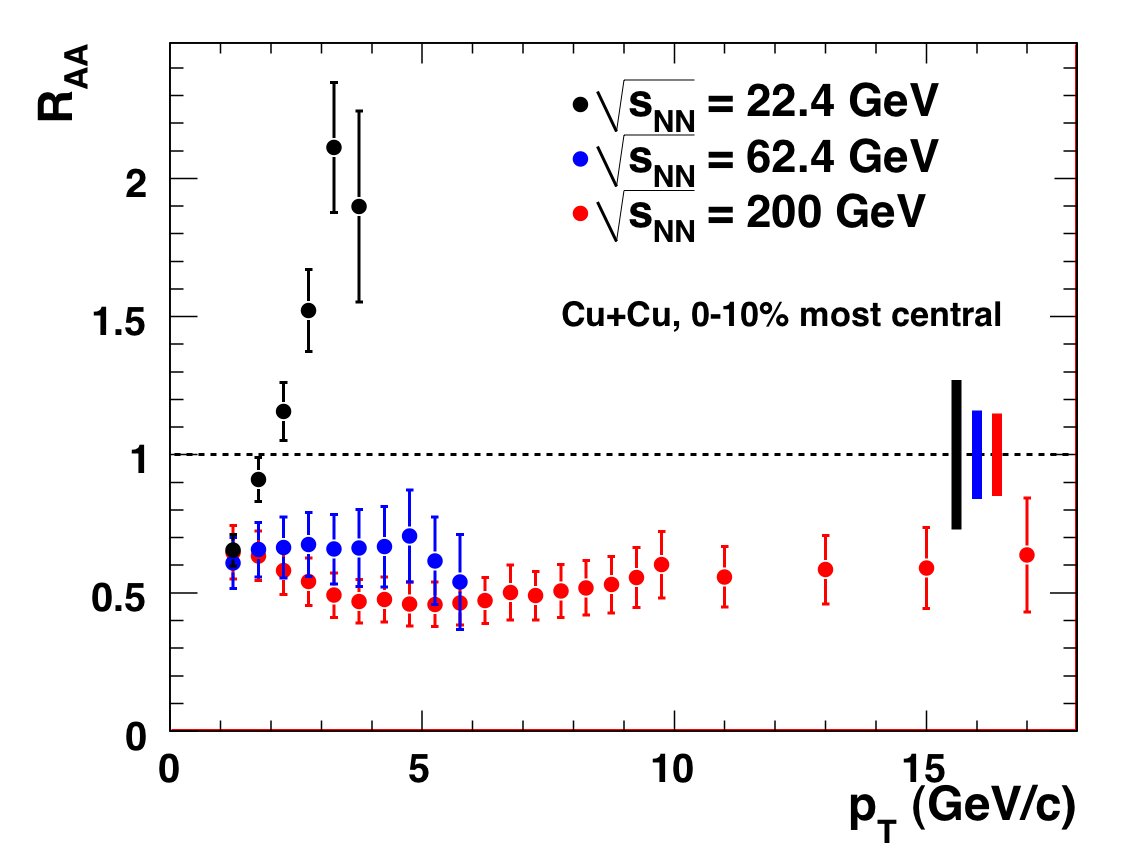}
    {\mbox{}\\\vspace{-0.3cm}
    \hspace{-0.2cm}\mbox{~} \bf (a)
    \hspace{+7.4cm}\mbox{~} \bf (b)}
    {\mbox{}\\\vspace{-0.2cm}}
    \caption
    {
      (a)~Higher order  moments of the net proton number measured for Au--Au collisions for different incident energies. 
      (b)~The nuclear modification factor of neutral pions as a function of transverse momentum for Cu--Cu
      collisions at three different energies.
      Figure (a) taken from \cite{Mohanty:2011nm} and figure (b) taken from \cite{Purschke:2011qm}.
    }
    \label{fig:STAR_PHENIX_bes2}
  \end{center}
\end{figure}
Figure \ref{fig:STAR_PHENIX_bes2}(a) shows results
for the higher order  moments (the skewness, $S$, and kurtosis, $\kappa$)
of the net proton number.
Small deviation of the conserved quantity, the baryon number,
from the Hadron Gas Resonance (HGR) model below 39 GeV/nucleon
may suggest a hint of proximity to the critical point.
Figure \ref{fig:STAR_PHENIX_bes2}(b) shows the charged particle $R_{\rm AA}$
down to an energy of 22.4 GeV/nucleon and it illustrates the significant change
in pattern of the particle suppression at lower energy compare to the top RHIC energy.
Overall, there are some hints from the Beam Energy Scan program at RHIC
on the possible critical point between energies of 7 and 20 GeV/nucleon.
Hopefully the upcoming measurements from RHIC energy scan will
allow us to make a conclusive statement about existence of the QCD critical point.

\section{\label{Sec:LPV}Probes of local parity violation in strong interactions}
The strong magnetic field created in the interaction zone
of non-central relativistic heavy-ion collisions
may interact with the topologically non-trivial gluonic field configurations
of QCD such as instantons and sphalerons.
It is predicted \cite{Kharzeev:2007jp} that experimentally this may lead
to charge separation of hadrons produced in the collision
along the magnetic field, which itself is aligned perpendicular to the
reaction plane of the collision.
Since instanton and sphaleron configurations breaks the parity symmetry of QCD,
the measurement of charge separation provides a unique 
experimental test of how well the parity symmetry is preserved by the strong interaction
\cite{Voloshin:2004vk}.

\begin{figure}[htb]
  \begin{center}
    \includegraphics[width=.47\textwidth]{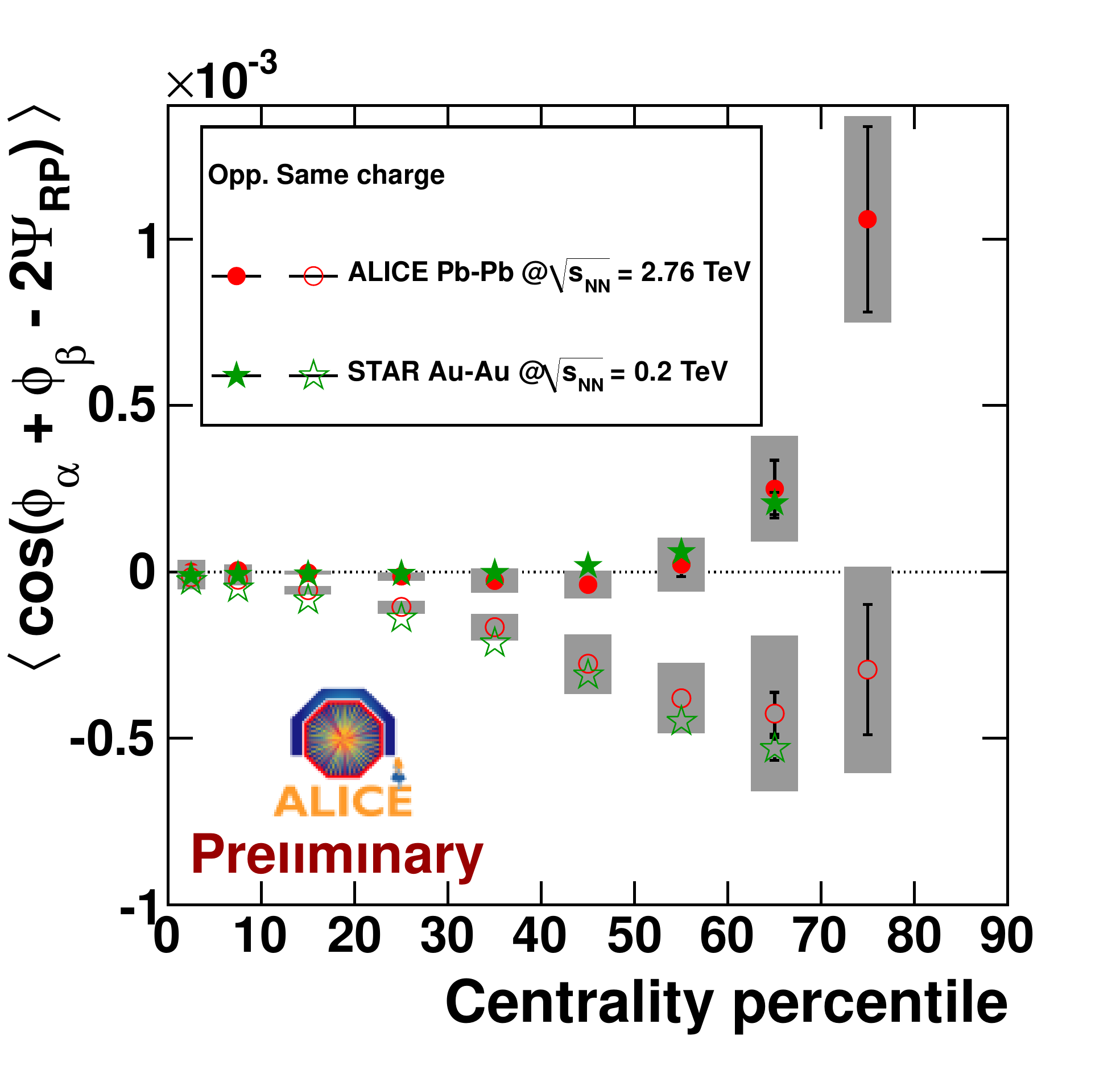}%
    \includegraphics[width=.51\textwidth]{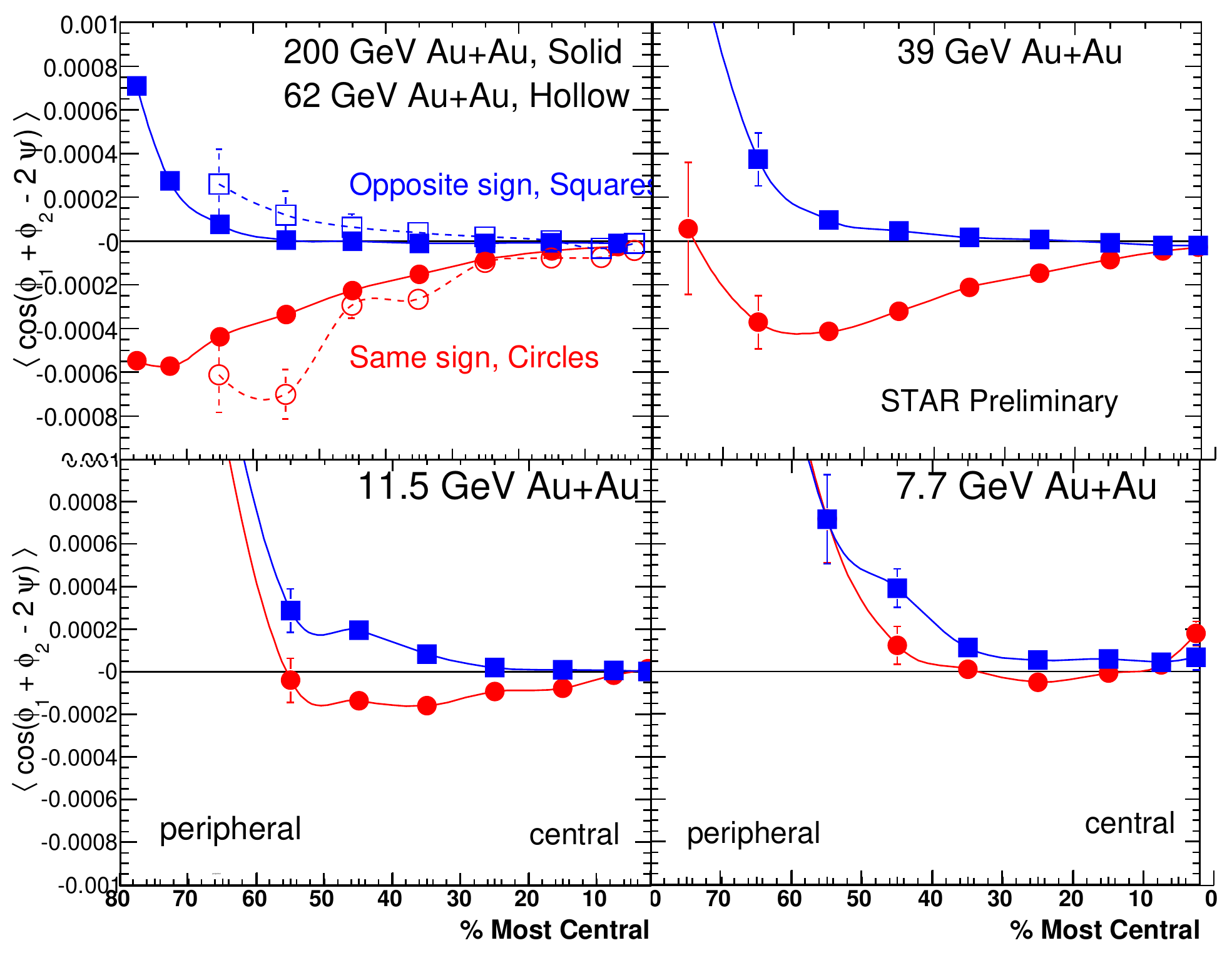}
    {\mbox{}\\\vspace{-0.2cm}
    \hspace{0cm}\mbox{~} \bf (a)
    \hspace{+6.7cm}\mbox{~} \bf (b)}
    {\mbox{}\\\vspace{-0.2cm}}
    \caption
    {
      Charged dependent azimuthal correlations with respect to the reaction plane of heavy-ion collision vs. centrality:
      (a)~comparison between results for Pb--Pb collisions at 2.76 TeV/nucleon vs. the measurement for Au--Au at 200 GeV/nucleon,
      (b)~results for Au--Au collisions for 5 different collision energies in the range of 7.7 - 200 GeV/nucleon.
      Note the inverted centrality scale in figures (a) and (b). 
      Figure~(a) taken from \cite{Collaboration:2011sm} and figure (b) from \cite{Mohanty:2011nm}.
    }
    \label{fig:ALICE_STAR_LPV}
  \end{center}
\end{figure}

Experimentally the effects of charge separation can be quantified  by the
charge dependent azimuthal correlations with respect to the reaction plane.
Figure \ref{fig:ALICE_STAR_LPV} shows the experimental results
for the $\langle \cos(\phi_{\alpha}+\phi_\beta- 2\Psi_{RP})\rangle$ correlator,
which is the parity even observable
but directly sensitive to the event-by-event charge fluctuations,
and thus to the possible local parity violation in strong interactions.
The STAR and PHENIX Experiments at RHIC, and now the ALICE Collaboration at LHC \cite{Collaboration:2011sm}
observe a significant charge separation
at higher collision energies (Fig.~\ref{fig:ALICE_STAR_LPV}(a)),
which seems to disappears between 11.5 and 7.7 GeV energies (lower panel in Fig.~\ref{fig:ALICE_STAR_LPV}(a)).
The experimental situation is significantly complicated by the
presence of the parity conserving background correlations
which may contribute to the measured charge dependent azimuthal correlations
at RHIC and LHC. Recent progress in understanding the possible
parity even backgrounds,
such as identifying the large flow fluctuations in the first harmonic flow
helps to better understand background contributions, but there
is still a long way to go before we will be able to conclude whether
the observed charge separation is indeed connected to
the effects of local parity violation, or it is just a complicated
interplay of yet unidentified background sources.

\section{Summary}

The results by the ALICE, ATLAS, and CMS Collaborations from the first heavy-ion run at
the Large Hadron Collider in November 2010
opened a new era of experimental studies
of the quark-gluon plasma  in the laboratory.
Together with the new high statistics data collected during the past few years
by the STAR and PHENIX Experiments at the Relativistic Heavy Ion Collider
this provides an extremely rich experimental data set
which allows us to study the properties of the quark-gluon plasma in the great detail,
and let us to learn more about the features
of the universe a few microseconds just after the Big Bang.
I am looking forward to further experimental developments and
more exciting results from the LHC and RHIC scientific communities!

\acknowledgements{%
%
This work was supported by the Helmholtz Alliance
Program of the Helmholtz Association, contract HA216/EMMI
"Extremes of Density and Temperature: Cosmic Matter in the Laboratory".
}


%

}  



\begin{thebibliography}{99}
  
\bibitem{Steinberg:2011dj}
  P.~Steinberg [ATLAS Collaboration],
  arXiv:1107.2182 [nucl-ex].

\bibitem{collaboration:2011ym}
  B. Wyslouch [CMS Collaboration],
  arXiv:1107.2895 [nucl-ex].

\bibitem{Floris:2011ru}
  M. Floris [ALICE Collaboration],
  arXiv:1108.3257 [hep-ex].

\bibitem{Aamodt:2011mr}
  K.~Aamodt {\it et al.} [ALICE Collaboration],
  Phys.\ Lett.\  {\bf B696 } (2011)  328-337.

\bibitem{Collaboration:2011yb}
  R. Snellings [ALICE Collaboration],
  arXiv:1106.6284 [nucl-ex].

\bibitem{Adare:2011tg}
  A.~Adare {\it et al.} [PHENIX Collaboration],
  arXiv:1105.3928 [nucl-ex].

\bibitem{Sharma:2011qm}
  D. Sharma [PHENIX Collaboration],
  Proceeding of Quark Matter 2011 Conference.


\bibitem{MartinezGarcia:2011nf}
  G.~Martinez Garcia [ALICE Collaboration],
    arXiv:1106.5889 [nucl-ex].

\bibitem{Andronic:2011yq}
  A.~Andronic, P.~Braun-Munzinger, K.~Redlich, J.~Stachel,
  arXiv:1106.6321 [nucl-th].

\bibitem{Lee:2011cs}
  Y.~-J.~Lee [CMS Collaboration],
  arXiv:1107.2131 [hep-ex].

\bibitem{Mohanty:2011nm}
  B.~Mohanty [STAR Collaboration],
  arXiv:1106.5902 [nucl-ex].

\bibitem{Purschke:2011qm}
  M.~L.~Purschke [PHENIX Collaboration],
  Proceeding of Quark Matter 2011 Conference.


\bibitem{Kharzeev:2007jp}
  D.~E.~Kharzeev, L.~D.~McLerran and H.~J.~Warringa,
  Nucl.\ Phys.\  A {\bf 803}, 227 (2008).

\bibitem{Voloshin:2004vk}
  S.~A.~Voloshin,
  Phys.\ Rev.\  {\bf C70}, 057901 (2004).


\bibitem{Collaboration:2011sm}
  P.~Christakoglou [ALICE Collaboration],
  arXiv:1106.2826 [nucl-ex].



\end{thebibliography}
\end{document}